\documentclass[12pt,aps,preprint,preprintnumbers,nofootinbib]{revtex4-1}
\usepackage{graphicx}
\usepackage{verbatim}
\usepackage{cancel}
\usepackage{amssymb}
\usepackage{amsmath,dsfont,textcomp}
\usepackage{bm}
\usepackage{color}
\usepackage{euscript,amsmath}

\def\beq{\begin{equation}}
\def\eeq{\end{equation}}
\def\bea{\begin{eqnarray}}
\def\eea{\end{eqnarray}}
\def\to{\rightarrow}

\newcommand{\tw}{\theta_{W}}

\begin{document}
\preprint{IPMU11-0050}
\title{A Revisit to Top Quark Forward-Backward Asymmetry}
\author{Jing Shu$~^{a,b}$}
\email{jing.shu@ipmu.jp}
\author{Kai Wang$~^{a,b}$}
\email{kai.wang@ipmu.jp}
\author{Guohuai Zhu$~^{a}$}
\email{zhugh@zju.edu.cn}
\affiliation{
$~^{a}$~Zhejiang Institute for Modern Physics~(ZIMP), Zhejiang University, Hangzhou, Zhejiang 310027, CHINA\\
$~^{b}$~Institute for the Physics and Mathematics of the Universe~(IPMU), the University of Tokyo, Kashiwa, Chiba 277-8568, JAPAN
}

\begin{abstract}
We analyze various models 
for the top quark forward-backward asymmetry ($A^t_{FB}$) at the
Tevatron, using the latest CDF measurements on different $A^t_{FB}$s and the total cross section. The axigluon model in Ref. \cite{paul} has difficulties in explaining the large rapidity dependent asymmetry and mass dependent asymmetry
simultaneously and the parameter space relevant to $A^t_{FB}$ is ruled out by the latest dijet search at ATLAS. In contrast to Ref. \cite{cp}, we demonstrate that the large parameter space in this model with a $U(1)_d$ flavor symemtry is not ruled out by flavor physics. The $t$-channel flavor-violating $Z^{\prime}$, $W^{\prime}$ and diquark models all have parameter 
regions that satisfy different $A_{FB}$ measurements within 1~$\sigma$. However, 
the heavy $Z^{\prime}$ model which can be marginally consistent with the total cross section is severely  constrained by the Tevatron direct search of same-sign top quark pair. The diquark model suffers from too large total cross section and is difficult to fit the $t \bar{t}$ invariant mass distribution. The electroweak precision constraints on the $W'$ model based on $Z'$-$Z$ mixings is estimated and the result is rather weak ($m_{Z'} > 450$ GeV). Therefore, the heavy $W^{\prime}$ model seems to give the best fit for all the measurements. The $W^{\prime}$ model predicts the $t\bar{t}+j$ signal 
from $tW^{\prime}$ production and is 10\%-50\% of SM $t\bar{t}$ at the 7 TeV LHC. Such $t+j$ resonance can serve as the 
direct test of the $W^{\prime}$ model.

\end{abstract}

\maketitle

The prompt decay of top quark before hadronization 
provides opportunity to explore its various properties
like charge, mass and spin. 
Given its large mass, the scale of top quark pair production is 
greater than $2m_{t}$ where the perturbative QCD plays important role. 
Therefore, the top quark pair production at hadron colliders 
can serve as handle of precision test of the standard model (SM) gauge
interaction, both  weak interaction in its decay and the perturbative QCD theory
of strong interaction in its production.
  
From the structure of the SM, top quark is special. As a colored particle, it is the heaviest known particle which is copiously produced at the hadron collider. Since the top quark acquire its large mass through the electroweak symmetry breaking (EWSB), any of its properties deviated from SM would be an important signals for new physics and potentially indicate the origin of EWSB, 
which makes searching new physics in the top quark sector 
extremely interesting at both Tevatron and LHC.

One important measurement for top quark in top quark pair production is the top forward backward asymmetry, which is equivalent to charge asymmetry under 
CP transformation \cite{qcd}. For the SM production, it involves the high precision calculation of QCD. At ${\cal O}(\alpha^{3}_{s})$, the bremsstrahlung amplitudes
$q\bar{q}\to Q\bar{Q}g$ carry an odd power of color charge
hence have an odd charge conjugation parity in the interference terms
among initial states radiation and final state radiation diagrams. There is also interference between the box diagram of ${\cal O}(\alpha^{4}_{s})$
with the LO diagram that contributes to the charge asymmetry.
 
CDF collaboration has recently updated  the measurements on the total 
 forward-backward asymmetry in top quark pair production 
 with the semi-leptonic $t\bar{t}$ data with integrated luminosity of 5.3~fb$^{-1}$
  \footnote{The top quark forward-backward asymmetry has also been 
 measured in the di-lepton channel as
$A_{FB} = 0.42 \pm 0.15(stat) \pm 0.05(syst)$
 in the $t\bar{t}$ rest frame with 5.1~fb$^{-1}$ data \cite{cdfdilepton}.}. 
  The observed total asymmetry measured in the lab frame and  the $t\bar{t}$ rest frame are
 \bea
A_{FB}^t &=& 0.150 \pm 0.050(\textrm{stat}) \pm 0.024(\textrm{syst}) ~~~ (p \bar{p} ~ \textrm{rest frame}) \nonumber \\
A_{FB}^t &=& 0.158 \pm 0.072(\textrm{stat}) \pm 0.017(\textrm{syst}) ~~~ (t \bar{t} ~ \textrm{rest frame})
\eea
which corresponds to the SM prediction based on the NLO simulation, 
Monte Carlo for FeMtobarn processes (MCFM), $0.038\pm0.006$ (in lab) and
 $0.058\pm0.009$ in $t\bar{t}$ rest frame respectively \cite{newcdf}.  
These measurements have improved the previous results based on 3.2~fb$^{-1}$ 
of $A^{p\bar{p}}_{FB}(\cos\theta)=0.19\pm 0.069$ and $A^{t\bar{t}}_{FB}(\Delta\eta)=0.24\pm 0.014$.
 \cite{newcdf} \footnote{Note that the recent D0 measurement $A_{FB}^t = (8 \pm 4\textrm{(stat)} \pm 1\textrm{(syst)}) \%$ is based on top-pair events that satisfy the experimental acceptance, which is uncorrected for effects from reconstruction or selection and can not be used to compare with the CDF results \cite{newd0}.} 

More importantly, with the enlarged data sample, 
CDF collaboration has also released two distributional measurements. 
The most interesting result is the mass dependent forward backward 
asymmetry. 
The mass dependent forward backward asymmetry in the $t\bar{t}$
rest frame
\beq
A^{t\bar{t}}_{FB}(M_{t\bar{t}}> 450~\text{GeV})=0.475\pm 0.112
\eeq
in comparison to the QCD correction prediction $0.088\pm 0.013$.
This $3.5~\sigma$ deviation may be a strong indication for physics
beyond the SM.
The second measurement is the rapidity dependent asymmetry, which is frame independent, as
\bea
A_{FB} (\mid \Delta y \mid > 1.0) &= &0.611 \pm 0.256 \\
A_{FB} (\mid \Delta y \mid < 1.0) &= &0.026 \pm 0.104\pm 0.056
\eea
in comparison to the MCFM SM prediction as $A_{FB} (\mid \Delta y \mid > 1.0)=0.123\pm 0.018 $ and 
$A_{FB} (\mid \Delta y \mid < 1.0) =0.039\pm 0.006$.

The ratio of the parton level asymmetries in the two different frames, which differ by longitudinal boost, is
\beq
{A^{p\bar{p}}\over A^{t\bar{t}}} = 0.95\pm 0.41
\eeq
with the error corrected for the expected correlation across frames in the NLO QCD assumption.
Even though the uncertainty is still large, this close to 1 central value 
implies that the top events which contribute to the asymmetry mostly
lie in the forward-backward direction so the asymmetries
are less dependent of the longitudinal boosts along the beam direction.
This feature is also shown in the $\Delta \eta$ dependent 
asymmetry $A^{t\bar{t}}_{FB} (\mid \Delta y \mid > 1.0)= 0.611$
which shows that the asymmetric events are mostly
due to events with larger rapidity difference $\mid \eta_{t}-\eta_{\bar{t}}\mid$.

On the other hand, the measurement of $t\bar{t}$ cross section $\sigma_{t\bar{t}}$, updated by the
4.6 fb$^{-1}$ CDF result (with $m_{t}=172.5$~GeV), is
$\sigma^\text{exp}_{t\bar{t}}=7.50 \pm 0.31 \text{(stat)} \pm 0.34 \text{(syst)} \pm 0.15 \text{(Z theory)}$~pb
which is  in very good agreement with SM theory prediction of $\sigma^\text{th}_{t\bar{t}}=7.5^{+0.5}_{-0.7}$~pb 
at NNLO \footnote{
The latest NNLL calculation shows the $\sigma_{t\bar{t}}(m_{t}=173.1~\text{GeV})= 6.30\pm 0.19^{+0.31}_{-0.23}~\text{pb}$ \cite{qcd} which is significantly lower than the experimental results. However, we still use the old SM predictions since we do not know $\sigma_{t\bar{t}}(m_{t}=172.5~\text{GeV})$ for the latest results. }.
Therefore, in order for new physics to generate large asymmetry without changing 
the total production cross section, the new physics contribution must interfere with  
the leading SM production of $u\bar{u},d\bar{d}\overset{g}\rightarrow t\bar{t}$ 
as color octet exchange in $s$-channel. For instance, in order for the  $s$-channel massive $Z'$ 
to explain the asymmetry, there is no interference between 
$s$-channel color singlet exchange $u\bar{u},d\bar{d}\overset{Z^{\prime}}\rightarrow t\bar{t}$ and QCD $u\bar{u},d\bar{d}\overset{g}\rightarrow t\bar{t}$. The asymmetry events due to $Z'$ 
will significantly enhance the total cross section $\sigma_{t\bar{t}}$ at the
same time and this causes a strong tension between fitting of $A_{FB}$ and $\sigma_{t\bar{t}}$.
This requirement implies that there are only two categories of candidate models
to solve this anomaly.
\begin{itemize}
\item  First category of models contain $s$-channel color octet vector boson but with parity violation at
both $q-\bar{q}-G$ and $t-\bar{t}-G$ vertices \cite{paul,schannel,baiyang,cp}.  

\item Second category correspond to the $t$-channel  exchange of light gauge boson of
maximal flavor violation that couples initial state $u,d$ quark to the third generation $t$ quark.
The large asymmetry can be generated via Rutherford singularity behavior \cite{hitoshi,waiyee,tim,tchannelz,james,tchannelw,tchanneldiquark,operator}. 
\end{itemize}

Both categories of models have their realizations in the beyond SM models.
Given the updated measurements, especially the new distributional measurements, 
we discuss the current status of various models. 
In addition, the models may have other implications that have been or will be 
constrained by some direct or indirect experiments. One realization
of the first category models is the non-universal axigluon model proposed in \cite{paul}
and it may receive constrain from low energy neutral meson mixings \cite{cp}. However, we show that the flavor bound can be easily evaded by putting a horizontal flavor symmetry $U(1)_d$.
In the $W_R$ models \cite{waiyee,tchannelw}, since the $W_R$ is charged under SM $U(1)_{em}$, the neutral component $W^3_R$ would inevitably mix with $W_L^3$ and some extra $U(1)_X$ which induce a $Z$-$Z'$ mixing. The new ATLAS Dijets \cite{dijet} search and the Tevatron same sign dileptons \cite{Aaltonen:2008hx} would severely constrain the s-channel axigluon models and the t-channel heavy $Z'$ models. We also study the direct prediction at the Large Hadron Collider(LHC) using the 1~$\sigma$ fitting of all three asymmetry measurements $A^{t\bar{t}}_{FB}(M_{t\bar{t}}> 450~\text{GeV})$, $A_{FB}^{t\bar{t}} (| \Delta y | > 1.0)$, $A^{t\bar{t}}_{FB}$(total) with the right total cross section.  

The paper is organized as follows. In Section \ref{sec:status}, we presented the 1~$\sigma$ fitting of all three asymmetry measurements for s-channel color octet model (Section \ref{sec:s-channel}), t-channel $Z'$ model (Section \ref{sec:t-channelzp}), $W'$ model (Section \ref{sec:t-channelwp}), diquark model (Section \ref{sec:t-channeldq}) and the corresponding consequences. In Section \ref{sec:LHC}, we calculate the production rates for the new particles in various different models at the Tevatron which give the bounds for those models and the LHC signals. In Section \ref{sec:indirect}, we consider some indirect bounds for the axigluon from flavor physics (Section \ref{sec:flavor}) and $W'$ model from electroweak precision test (EWPT) (Section \ref{sec:ewpt}). Section \ref{sec:conclusion} contains our conclusions.

\section{Updated status of the models}
\label{sec:status}
In this section, we discuss the updated status of the models based on
the latest measurements, especially the new distributional measurements.
The simulation in the following discussion is at parton level and leading order.
The asymmetry observables are defined at parton level without taking
into account possible reconstruction efficiency. The SM contribution to the 
asymmetries from MCFM simulation have been subtracted to the corresponding 
measured values. 
The total cross section is obtained by multiplying a QCD $k$-factor.  
Since the latest experimental value is based on $m_{t}=172.5$~GeV,
for better comparison, we employ the theory calculation at NNLO
for $m_{t}=172.5$~GeV and the $k$-factor is 1.3.
Last, the differential cross section of $t\bar{t}$ invariant mass is not included as
requirement in the scan since QCD correction \cite{zhushouhua} and cut efficiency \cite{kathryn} 
may significantly modify the shape of differential distribution $d\sigma/d_{M_{t\bar{t}}}$.

In the following discussion, we are mostly interested in the region where the three asymmetry measurements can be explained within 1~$\sigma$. 

\subsection{$s$-channel color octet}
\label{sec:s-channel}

The interference term between the color octet $V-A$ gauge boson $G^{a}_{\mu}$ contribution and gluon
contribution in $q\bar{q}$ annihilation pickup a term as
\beq
\frac{2 g^{2}_{s}\hat{s} (\hat{s}-M_G^2)}
{(\hat{s}-M_G^2)^2+M_G^2 \Gamma_G^2}
\left[+ 2 \, g_A^q \, g_A^t \, \beta\cos\theta  \right]
\eeq
where $g_{s}$ is the strong coupling, $g^{q}_{A}$ is the axial component of the coupling between $G^{a}_{\mu}$
and light quarks $q$ and $g^{t}_{A}$ is that of the top quark. If the interference contribution is
positive asymmetry, it requires that the axial coupling $g^{q}_{A}g^{t}_{A}<0$  is 
inevitable \footnote{This non-universal gauge interaction potentially cause the violation of GIM mechanism thus
may be constrained from flavor changing neutral current (FCNC) processes such as neutral meson mixings
and we discuss its implications in the next section.}.

The $s$-channel models can be realized in various context. 
The first realization is the axigluon models 
where $SU(3)_{c}$ color gauge symmetry is only 
a remnant of $SU(3)_{L}\times SU(3)_{R}$ broken by a 
bi-triplet scalar and another color octet with axial coupling
become massive. However, to achieve the $g^{q}_{A}g^{t}_{A}<0$ requirement, 
the axigluon model has to be non-universal and one example
is the 4-generation model proposed in \cite{paul}.
Another realization is the models of extra dimension theory
where massive color octet Kaluza-Klein (KK) gluon couple to the SM quarks
in the chiral form as a result of fermion profiles \cite{ued}. The large 
$m_{t}$ naturally implies that the top quark 
and light quarks couple to $KK$ gluon in different way. 

One interesting feature that was discussed in the 
axigluon model \cite{paul} is the mass dependent
asymmetry. Due to the opposite contribution to
asymmetry between the interference term and
new physics squared term, the asymmetry is positive
when the centre-of-mass energy is 
at intermediate energy but when it is
close to the threshold, the asymmetry
may become negative. This bending-over in correlation
between asymmetry $A^{t}$ and the centre-of-mass energy 
$M_{t\bar{t}}$ had been shown in the latest CDF measurements,
in both the measurement with finite bin sizes of $M_{t\bar{t}}$
and the measurement with below/above $M_{t\bar{t}}$ edge. 

We use the axigluon model as one example to illustrate the
feature of $s$-channel models in comparison with 
the updated measurement.
\begin{figure}[htbp]
\begin{center}
\includegraphics[angle=0,width=8cm]{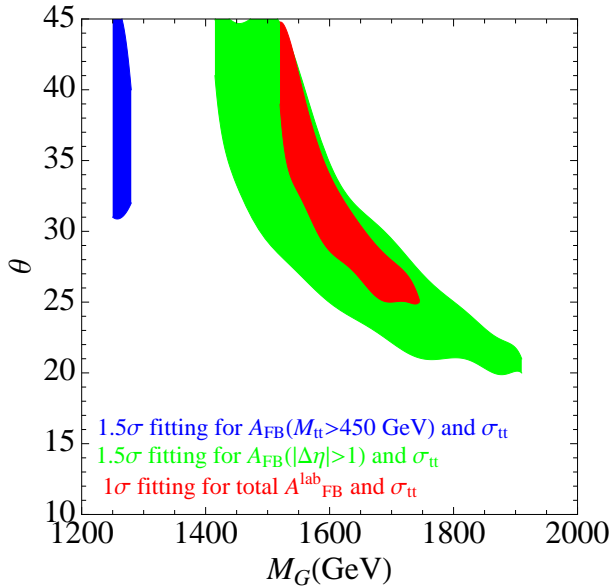}   
\caption{1~$\sigma$ parameter region for constraints from the total asymmetry in the lab frame
and the $t\bar{t}$ production rate; 1.5~$\sigma$ parameter region for the $A_{FB}$ in
$M_{t\bar{t}}>450$ and $\mid\Delta\eta\mid>1$ and $\sigma(p\bar{p}\to t\bar{t})$.   
}
\label{axigluon}
\end{center}
\end{figure}
Figure \ref{axigluon} shows the summary of best fit parameter
regions for total asymmetry, mass dependent asymmetry,
rapidity dependent asymmetry, the total cross section and
the last bin of $d\sigma /d M_{t\bar{t}}$ measurements.
Since the total asymmetry has been reduced from the
previous fitting in \cite{paul}, the 1~$\sigma$ region 
with total asymmetry is enlarged as shown in Fig.\ref{axigluon}.
However, there is no 1~$\sigma$ region for 
the mass dependent asymmetry of $M_{t\bar{t}}>450$~GeV or the
rapidity dependent asymmetry of $\mid\Delta \eta\mid>1$. 
Figure \ref{axigluon} shows the 1.5~$\sigma$ parameter space for 
$A_{FB}(\mid\Delta \eta\mid>1)$, $A_{FB}(M_{t\bar{t}}>450~\text{GeV})$ as
well as the total $t\bar{t}$ production rate $\sigma(p\bar{p}\to t\bar{t})$. 
It is clearly shown that the axigluon model \cite{paul} does
not consistently generate the large asymmetries in the events 
of $M_{t\bar{t}}>450$~GeV and $\mid\Delta \eta\mid>1$ \footnote{One can use the
general color octet vector boson with $V-A$ interaction to fit the two distributional
asymmetries and total asymmetry \cite{newcdf}. The general results together with the most recent bounds in the dijet channel from ATLAS will be presented elsewhere.}.

\subsection{$t$-channel}

As we argued,  the ratio of $A^{p\bar{p}}/A^{t\bar{t}}$
close to one may imply that the top events are mostly
in the forward-backward direction so the asymmetries
are less dependent of the 
longitudinal boosts along the beam direction.
Since the $t$-channel models naturally predict large number 
of events in the forward-backward region, 
the close to one ratio of $A^{p\bar{p}}/A^{t\bar{t}}$ is a 
basic feature of $t$-channel models. 

If the asymmetry is due to new physics in $t$-channel physics,
the interference contribution between new physics and SM QCD is proportional to 
\beq
C_{F} \frac{g^{2}_{s} g^{2}_{NP}}{\hat{s} t_{t}} (u^{2}_{t}+\hat{s} m^{2}_{t}+...),
\eeq  
where $t_{t} = -{1\over 2}\hat{s}(1-\beta\cos\theta)$ and $1/t$ expansion naturally picks up 
a $\cos\theta$. The $t$-channel physics naturally generates
a large asymmetry in the $t\bar{t}$ system. In addition,
the maximal asymmetry is generate at the Rutherford singularity
where $\theta=0$ which corresponds to very high centre-of-mass
energy. One would then expect the positive correlation between
$A^{t}_{FB}$ and $M_{t\bar{t}}$.  

The $t$-channel $Z'$ model in \cite{hitoshi,tchannelz} 
proposed a color singlet neutral gauge boson
with maximal flavor violation between first and third
generations and the new contribution interferes 
with the SM $u\bar{u}\to t\bar{t}$. Similar to the $Z'$ model, 
instead of neutral current exchange in $t$-channel,
there is also a proposal using charge current exchange
in $t$-channel as flavor violation $W^{\prime}$.
The interference effect is reduced since it's only
the $d\bar{d}$ initial state \cite{waiyee,tchannelw}. 
Such flavor violation gauge interactions may be
realized in horizontal gauge interaction models \cite{james} 
for neutral current or generalized left-right model \cite{tchannelw} for 
charged current.
 
A Higgs-like scalar with maximal flavor violation\cite{rajaraman} 
would generate a large negative asymmetry
due to the helicity-flip in the Yukawa coupling.
The spin conservation in the $\theta=0$ direction
requires the top quark to move backward. 
To resolve this, the fermion-number violating diquark scalars 
with maximal flavor violation was proposed \cite{tim,tchanneldiquark,wise}. 
Diquark scalar can be  $3\otimes 3= 6\oplus\bar{3}$ under $SU(3)_{c}$
and has fermion-number violating coupling 
as $\overline{t^{c}}u \phi$ or $\overline{t^{c}} d \phi$.  
Such diquark scalars with flavor violation can also be realized in 
various BSM contexts, partial unification models or supersymmetry. 
For instance, $R$-parity violation supersymmetric standard model 
which contains the baryon number violating coupling, $\epsilon_{\alpha\beta\gamma}u_{\alpha}^{c}d_{\beta}^{c}d_{\gamma}^{c}$ \cite{rparity}, the down type squark $\tilde{d}_{i}$
can mediated $u$-channel $d\bar{d}\to t\bar{t}$ that interferes with
the QCD $d\bar{d}\to t\bar{t}$.

All the three proposals can in principle predict large
positive asymmetry in $t\bar{t}$ production. In the following
paragraphs, we examine the numerics to see whether 
the models can explain the three asymmetry measurements
and the total cross section at the same time.

\subsubsection{$Z'$}
\label{sec:t-channelzp}

\begin{figure}[htbp]
\begin{center}
\includegraphics[angle=0,width=8cm]{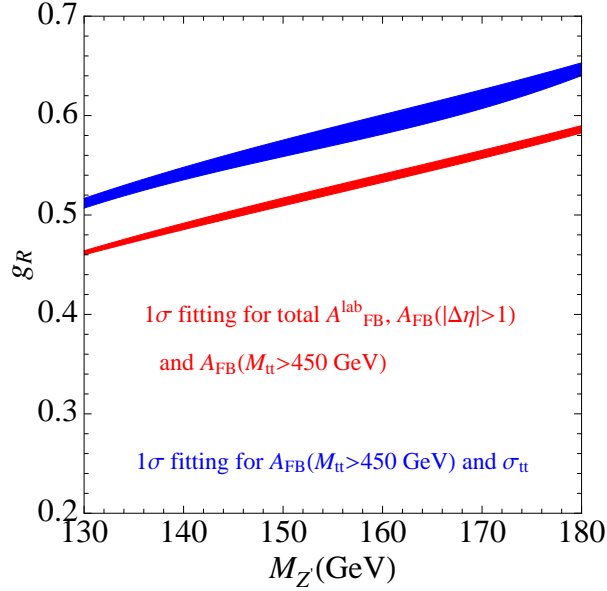}   
\caption{Parameter space scan of 1~$\sigma$ for all three asymmetries, $A^{total}_{FB}$,
$A_{FB}(\mid \Delta\eta\mid>1)$ and $A_{FB}(M_{t\bar{t}}>450~\text{GeV})$ is shown in
red. The 1~$\sigma$ fitting for $A_{FB}(M_{t\bar{t}}>450~\text{GeV})$ and the total cross section 
$\sigma_{t\bar{t}}$.}
\label{zprime}
\end{center}
\end{figure}

We first examine the first proposed $t$-channel model, $Z'$ \cite{hitoshi}. 
To minimize the constraints from low energy, the authors proposed
a right-handed coupled $Z'$ with large coupling between $u$ and $t$. The parameter region
for 1~$\sigma$ fitting of all three asymmetry 
measurements as well as the total cross section for 
light $Z'$ mostly below $t$ threshold is presented in Figure \ref{zprime}. Due to large destructive interference,
the total cross section is always \textsl{smaller} than the
measured value. This result is also shown
in the NLO calculation of $Z'$ model \cite{zhushouhua}.
The best fit points for heavy $Z'$ by requiring 
1~$\sigma$ fitting for all the three asymmetry
measurements are listed in the Table \ref{1sigmazprime}. 
The corresponding $t\bar{t}$ cross section are also
below the 1~$\sigma$ total cross section and the
best points are towards heavy masses of ${\cal O}$(700~GeV).

\begin{table}[htdp]
\begin{center}
\begin{tabular}{|c||c|c|c|c|}
\hline
$M_{Z^{\prime}}$, $g_R$ & $A^{total}_{FB}$ & $A_{FB} (M_{t\bar{t}}>450~\text{GeV})$ & $A_{FB}(|\Delta\eta|>1)$  & $\sigma_{t\bar{t}}$  (pb) \\
\hline
275, 0.8 & 15.4\% & 32.7\% & 23.5\% & 6.4\\
450, 1.2 & 15.8\% & 34.4\% & 23.4\% & 6.6\\
575, 1.5 & 16.6\% & 35.9\% & 24.4\% & 6.8\\
700, 1.8 & 16.7\% & 36.1\% & 24.7\% & 6.9\\
750, 1.9 & 15.9\% & 34.7\% & 23.2\% & 6.9\\
\hline
CDF & 5.7\%--16.7\% & 27.5\%--50.0\% & 23.1\%-- 74.5\% & $7.5\pm 0.48$\\
\hline
\end{tabular}
\end{center}
\caption{1~$\sigma$ benchmark points for all three asymmetry measurements. $k$-factor = 1.3, $m_{t}$=172.5 GeV for $\sigma_{t\bar{t}}$. }
\label{1sigmazprime}
\end{table}%

One more complication which has been discussed
in \cite{hitoshi,waiyee,tim} is that the events in 
the $t$-channel exchange tend to be in high 
energy region which significantly increase the 
tail of $d \sigma/d M_{t\bar{t}}$, especially the last bin
(800~GeV--1.4~TeV) in $d \sigma/d M_{t\bar{t}}$.
The QCD correction may change the shape and lower 
the contribution in high energy \cite{zhushouhua}.
In addition, the $t$-channel kinematics implies that
the top quark events at high energy are mostly
in the larger rapidity region while the selection cut
are more efficient for the central events. Consequently,
the cut efficiency at high invariant mass 
is quite low \cite{kathryn}, which may further decrease the effective total cross section. 
Polarization of top quark in the events sample 
also effect the cut efficiency.

\subsubsection{$W'$}
\label{sec:t-channelwp}

\begin{figure}[htbp]
\begin{center}
\includegraphics[angle=0,width=8cm]{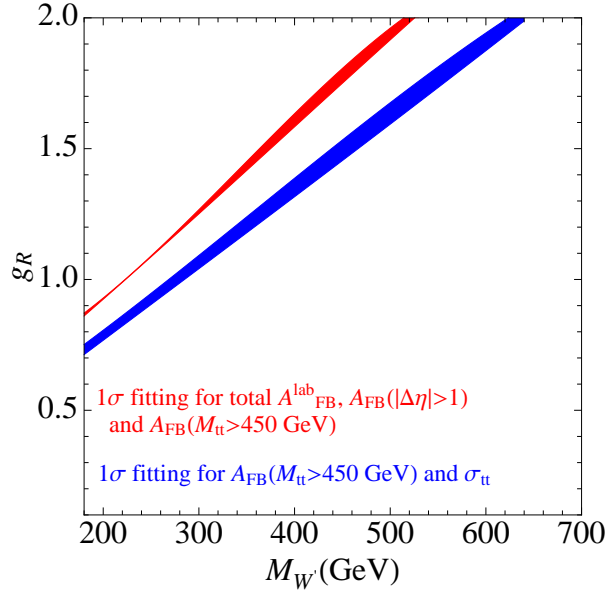}
\caption{Parameter space scan of 1~$\sigma$ for all three asymmetries, $A^{total}_{FB}$,
$A_{FB}(\mid \Delta\eta\mid>1)$ and $A_{FB}(M_{t\bar{t}}>450~\text{GeV})$ is shown in
red. The 1~$\sigma$ fitting for $A_{FB}(M_{t\bar{t}}>450~\text{GeV})$ and the total cross section 
$\sigma_{t\bar{t}}$.}
\label{totalw}
\end{center}
\end{figure}

To resolve the tension between cross section
and total asymmetry in the $Z'$ model, the charged
current process in $t$-channel may give better fit 
which has smaller interference effect due to the $d\bar{d}$ initial state. 
We plot the allowed parameter regions for the $t$-channel
charged current model in \cite{waiyee} in
Fig \ref{totalw}.

The 1~$\sigma$ asymmetry region of all three measurements 
corresponds to a larger total cross section which is outside
the 1~$\sigma$ fit of latest $\sigma_{t\bar{t}}$ measurement. 
However, various efficiency effects discussed in
the last paragraph of $Z^{\prime}$ session may 
significantly reduce the measured cross section. 

\subsubsection{Diquark}
\label{sec:t-channeldq}

We use the anti-triplet diquark that couples
to $\overline{t^{c}} u \phi$ to illustrate the feature.
Similar to the $W^{\prime}$ case, there also 
exist diquark scalars whose couplings are of $\overline{t^{c}} d \phi$
and these diquark scalars contribute to $d\bar{d}\to t\bar{t}$ instead.

\begin{figure}[htbp]
\begin{center}
\includegraphics[angle=0,width=8cm]{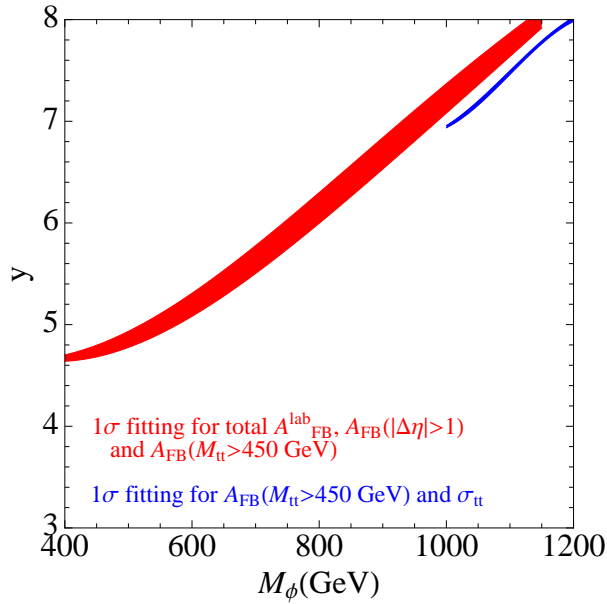}
\caption{Parameter space scan of 1~$\sigma$ for all three asymmetries, $A^{total}_{FB}$,
$A_{FB}(\mid \Delta\eta\mid>1)$ and $A_{FB}(M_{t\bar{t}}>450~\text{GeV})$ is shown in
red. The 1~$\sigma$ fitting for $A_{FB}(M_{t\bar{t}}>450~\text{GeV})$ and the total cross section 
$\sigma_{t\bar{t}}$.}
\label{diquark}
\end{center}
\end{figure}

Figure \ref{diquark} gives the 1~$\sigma$ fitting for
the anti-triplet diquark scalar with maximal flavor violation.  
The 1~$\sigma$ region also exists for the anti-triplet
diquark scalar for all the measurements in asymmetries $A^{t}_{FB}$.
But the corresponding total cross section $\sigma_{t\bar{t}}$
are also larger than the measured value over 1~$\sigma$. 
In addition, the best-fit for cross section and the mass dependent asymmetry 
is over 1~TeV which makes the $d\sigma /d M_{t\bar{t}}$ 
measurement very difficult to fit as shown in \cite{tim}.
The latest simulation by \cite{kathryn} also showed that
the $t\bar{t}$ events generated by diquark scalar 
had a higher cut efficiency at high energy therefore the
anti-triplet diquark fitting is worse than the $W^{\prime}$.

\section{Implications at the Tevatron and LHC}
\label{sec:LHC}
After fitting the top forward backward asymmetries in different kinematical regions, we discuss the other Tevatron bounds for the models and the LHC predictions that can be soon tested in this section. 

The Large Hadron Collider~(LHC) is a proton-proton collider with centre-of-mass energy
7~TeV in the first two years running. 
Unlike at Tevatron where the axigluon effect only appears as interference. 
The color octet axigluon of $\cal O$(1~TeV) can
be directly produced at the LHC and decay into dijet or $t\bar{t}$. With significant decay 
branching ratio (BR) to $t\bar{t}$, it provides additional handle to search it. 
The study of axigluon at the LHC has been performed by \cite{baiyang}. 
ATLAS collaboration has recently released the search for dijet resonance.
The latest data has ruled out axigluon from 0.6-2.1 TeV by assuming axigluon
coupling is only $g_{s}$. The axigluon model in \cite{paul} has a even larger 
coupling comparing with the ATLAS paper and therefore, the model receive
much more server constraint. 

For neutral gauge boson like $Z'$, the flavor violating vertex of $ut$ will lead to 
large $uu\to tt$ or $\bar{u}\bar{u}\to \bar{t}\bar{t}$ 
scattering with $Z'$ exchange in the $t$/$u$-channel. The same-sign positive top
quark pair ($uu\to tt$) becomes particular interesting at the LHC given its 
large $u$-valence quark parton flux \cite{uutt}. 

In addition, with large $ut$ coupling,
the $t Z^{\prime}$ associate production is not negligible. Since $Z^{\prime}$ equally
decays into $u\bar{t}$ and $t\bar{u}$, the associated production $t Z^{\prime}$ or $\bar{t} Z^{\prime}$
will contribute to $tt+j$, $\bar{t}\bar{t}+j$  and $t\bar{t}+j$ final states. 
Again, since the LHC is proton proton collider, the $tt+j$ dominates the 
same-sign top production. The $t\bar{t}+j$ will appear in the inclusive $t\bar{t}$ search. 
Since the 1~$\sigma$ parameter
space of all the asymmetry constraints corresponds to smaller $t\bar{t}$ pair
production, the additional $t\bar{t}+j$ may in principle help to ease the tension at Tevatron.
However, if it significantly contribute to the $t\bar{t}$, the same amount of
same-sign top quark will arise. 

\begin{figure}[htbp]
\begin{center}
\includegraphics[angle=0,width=8cm]{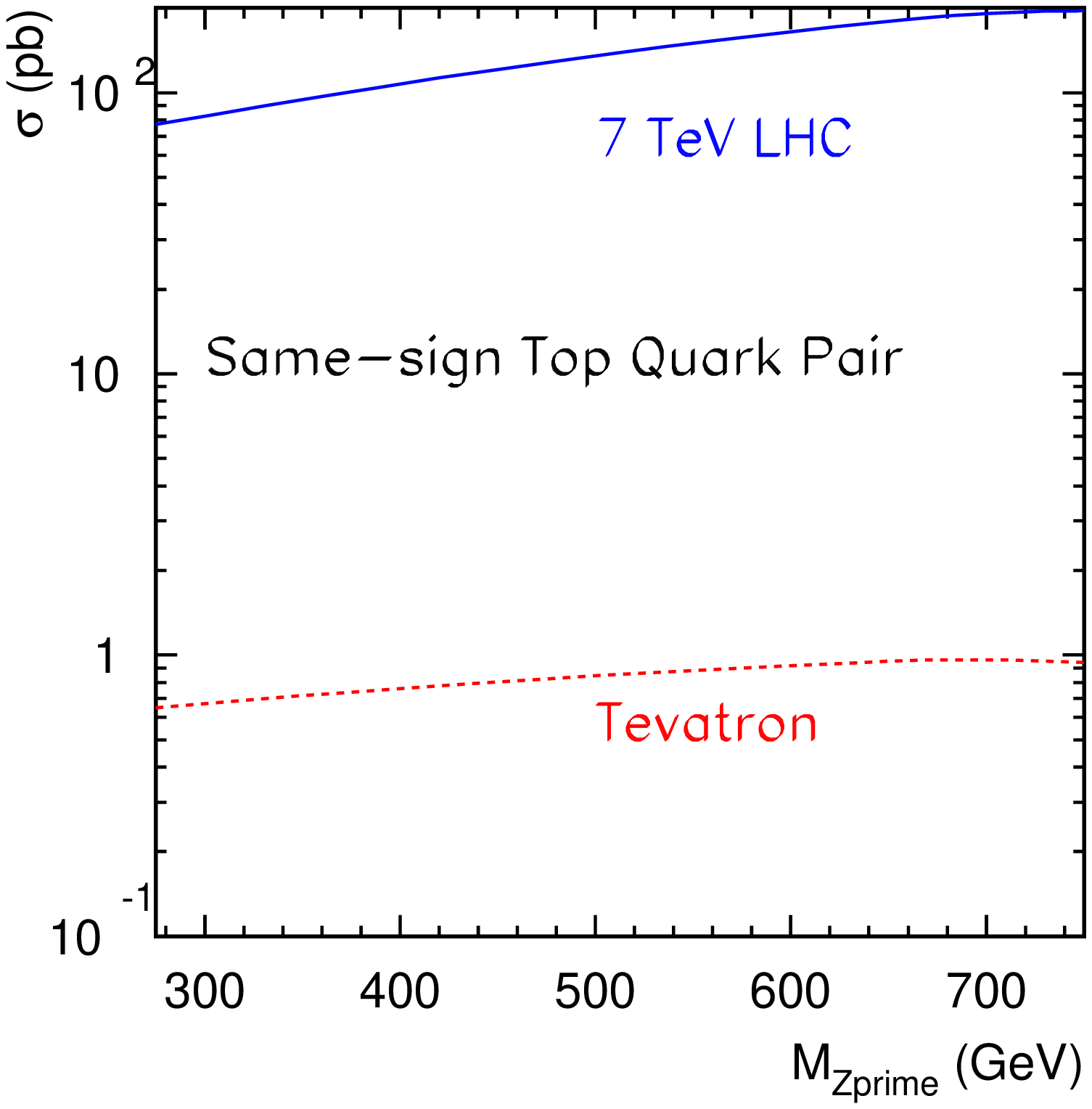}   
\includegraphics[angle=0,width=8cm]{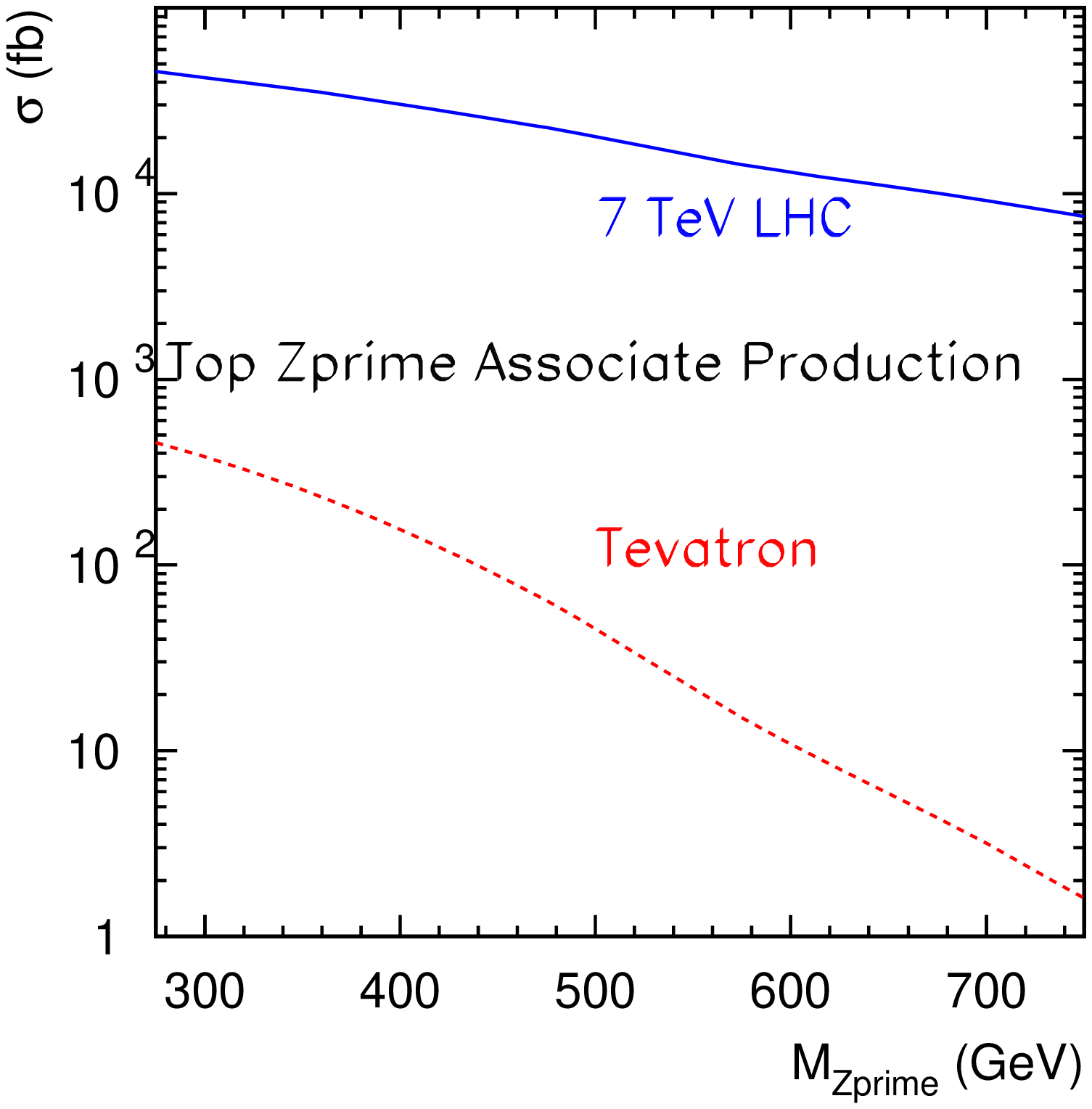}   
\caption{(a) $\sigma(pp\to tt)$ at Tevatron and the 7 TeV LHC; (b) $\sigma(pp\to t Z^{\prime}+\bar{t}+Z^{\prime}$) 
at Tevatron and the 7 TeV LHC. Both (a) and (b) are based on model parameter of $Z^{\prime}$ in the
1~$\sigma$ fitting of all three asymmetry measurements as listed in Table \ref{1sigmazprime}.}
\label{uutt}
\end{center}
\end{figure}

Figure \ref{uutt} (a) gives the $pp\to tt$ production rate at Tevatron and the 7 TeV LHC 
with the $Z^{\prime}$ in the 1~$\sigma$ fitting for all three asymmetry measurements. 
The $p\bar{p}\to tt+\bar{t}\bar{t}$ at Tevatron is between 0.7--1~pb for these best fit points. 
CDF measured only 3 events for 2~fb$^{-1}$ \cite{Aaltonen:2008hx} with the acceptance range from 1.5$\%$ to 3$\%$. The best fit points all predict 15-30 same-sign pure leptonic top events before selection cut but with one b-tagging. Even though these events from $t$-channel vector boson exchange may suffer from a low cut efficiency comparing to the $t$-channel \textsl{light} scalar exchange considered in Ref. \cite{Aaltonen:2008hx}, the $Z^{\prime}$ model is strongly 
constrained by the same-sign top quark scattering. 
At Tevatron, the same-sign top due to $t Z^{\prime}$ associate production is then much suppressed at the Tevatron due to significant phase space suppression. 

The $uu\to tt$ scattering get significantly enhanced at the proton-proton collider LHC. The 
production rate can reach 200 pb. Therefore, even at very early running of LHC with about
30~pb$^{-1}$ data and requiring two b-tagging jet, the event number before kinematic cut
is about 70 and the same-sign top quark $tt$ events is expected to be ${\cal O}(10)$.

For $W^{\prime}$ or diquark scalars with flavor violation, since they are electrically charged,
it will only contribute to $t\bar{t}$ as at Tevatron. However, since
the $W^{\prime}$ or diquark $\phi$ has a large $dt$ or $ut$ coupling,  the $d g\to t W^{\prime}$ or $ug\to \bar{t} \phi$
production is significant as shown in \cite{waiyee, tim,kathryn1}. With $W^{\prime}$ and
diquark scalars of typically above top quark threshold, they can decay into $t$ plus one hard jet. 
The signal is then $t\bar{t}$ plus one hard jet and should appear in the inclusive $t\bar{t}$ searches.
The diquark case has already been calculated in our early paper \cite{tim}.

\begin{figure}[htbp]
\begin{center}
\includegraphics[angle=0,width=8cm]{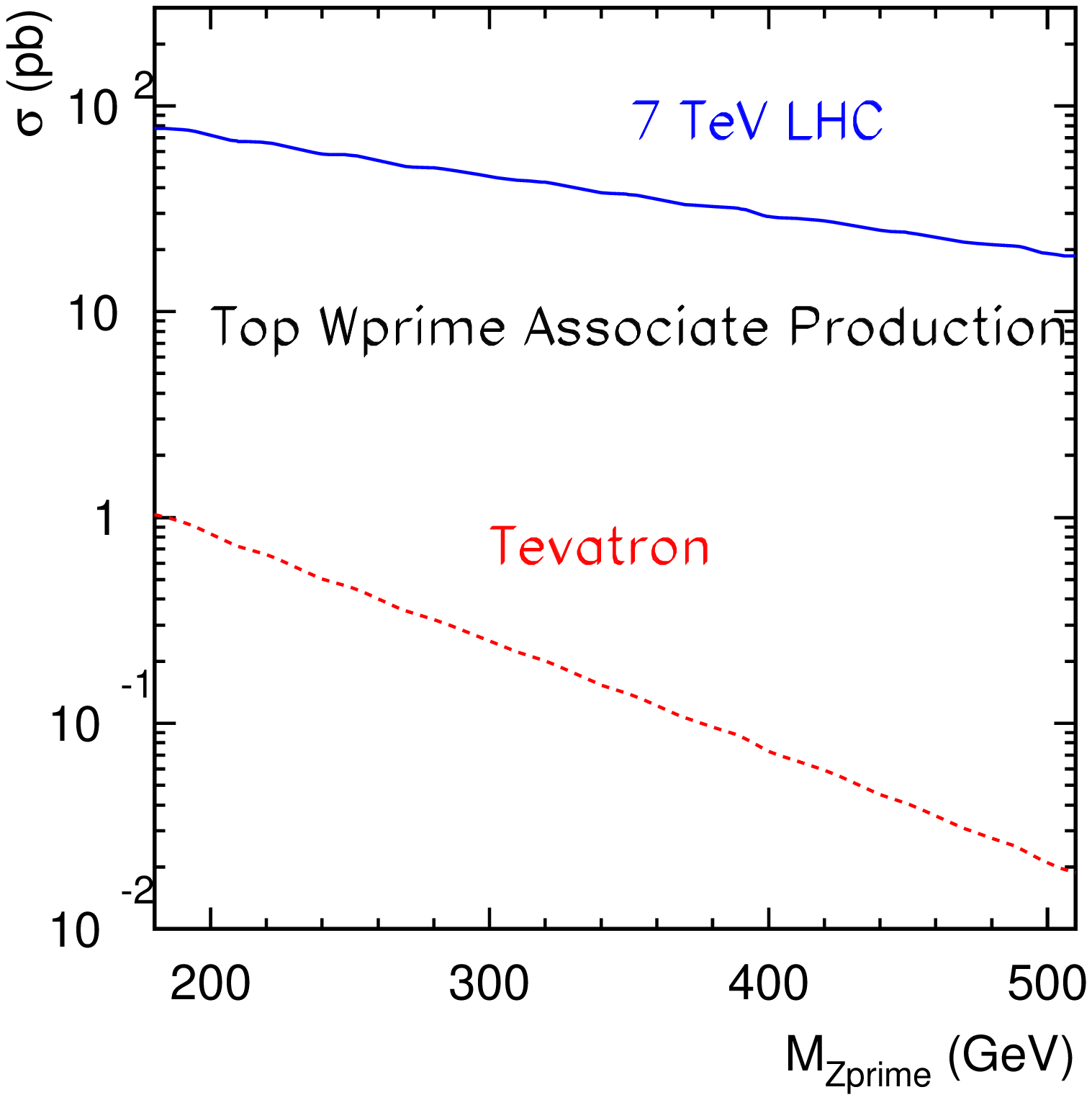}   
\caption{$pp\to tW^{\prime}\to j+t\bar{t}$}
\label{twp}
\end{center}
\end{figure}

Figure \ref{twp} gives the production of top quark plus $W^{\prime}$ at hadron colliders.  
For $M_{W^{\prime}}$ lighter than 400 GeV, the production rate
is about 0.1--1~pb at the Tevatron. As we discussed in the previous section, 
the 1~$\sigma$ fitting parameter space for all the asymmetries 
constraint corresponds to the larger cross section region. \footnote{\cite{Craig:2011an} claims that the single top production at Tevatron puts a strong constraint on the $W^{\prime}-u-b$ coupling. However this constraint does not apply to general $W^\prime$ models.}  
The new contribution to $t\bar{t}+j$ will increase the tension
between $A_{FB}$ and $\sigma_{t\bar{t}}$.  
For heavy $W^{\prime}$ above 400 GeV, due to large phase space 
suppression, the production rate at Tevatron can then be neglected. 
However, at the LHC, even with 7 TeV centre-of-mass energy, 
the production rate is ${\cal O}(10~\text{pb})$.  
Even though $gg\to t\bar{t}$ dominates the $t\bar{t}$ at the LHC which 
does not interfere with $W^{\prime}$, the $tW^{\prime}$ itself may 
significantly increase the $t\bar{t}$ rate. 

\section{Indirect constraints for models}
\label{sec:indirect}

\subsection{Axigluon with flavor protection}
\label{sec:flavor}
As shown in the previous section, only non-universal axigluon models can
provide the positive asymmetry. Being a color octet with strong coupling strength, 
this GIM violation axigluon will then lead to significant flavor changing neutral current (FCNC)
effect. 
\beq
{\mathcal L}= i g_{L}\bar{q}^{i} \gamma^{\mu} \left(H^{q}_{L}\right)_{ij} P_{L}q^{j} T^{a}G^{a}_{\mu}+i g_{R} \bar{q}^{i} \gamma^{\mu} \left(H^{q}_{R}\right)_{ij} P_{R}q^{j} T^{a} G^{a}_{\mu}
\eeq

Flavor violation thus can arise from the non-universal gauge couplings due to the rotation between mass eigenstate and gauge eigenstate.
\beq
u_{L} = V^{u}_{L} u_{L}, ~d_{L} = V^{d}_{L} d_{L}, ~u_{R} = V^{u}_{R} u_{R},~d_{R} = V^{d}_{R} d_{R}
\eeq

The effective coupling in horizontal space is
\beq
{V^{u}_{L}}^{\dagger} H^{u}_{L} V^{u}_{L}, {V^{u}_{R}}^{\dagger} H^{u}_{R} V^{u}_{R}, {V^{d}_{L}}^{\dagger} H^{d}_{L} V^{d}_{L}, {V^{d}_{R}}^{\dagger} H^{d}_{R} V^{d}_{R},
\eeq

The rotation from mass eigenstate to gauge eigenstate for up and down type quark respectively is completely unmeasurable in
the weak interaction. The only observable is the mixing in charge current transition
which is categorized as CKM matrix
\footnote{It was argued in \cite{cp} that the axigluon model in \cite{paul}
suffers serve bounds from $B_{d}$ mixing. However, the calculation seems 
to be done by assuming both up and down quark sectors transform like CKM 
rotation. }.

To avoid flavor violation in the down sector, one may introduce a $U(1)_d$ symmetry \cite{Csaki:2008eh}
which acts only on the down sector with different eigenvalues for different generations, but does not distinguish the handedness of the quarks. Then the down-quark sector are diagonal with 
$V^{d}_{L}=V^{d}_{R}=\mathds{1}$ so that there is no FCNC at all in the $B_{s}$, $B_{d}$ or neutral $K$ system.
For simplicity, we take further the rotation matrix of right-handed up-quark sector as $V^{u}_{R}=\mathds{1}$. 
Then one can explicitly determine the left-handed up-type quark rotation based on the known CKM matrix using $V^{u}_{L} {V^{d}_{L}}^{\dagger}=V_{CKM}$
\beq
V^{u}_{L}=V_{CKM}
\eeq

Nowadays the Wolfenstein parametrization \cite{Wolfenstein:1983yz} is widely used to express the CKM matrix in terms of four parameters
($\lambda$, $A$, $\rho$ and $\eta$). To keep the unitarity of CKM matrix to all orders of $\lambda$, we adopt in the following a definition of
Wolfenstein parameters proposed in \cite{Buras:1994ec}. Then the effective coupling between up quark and charm quark is
\beq
({V^{u}_{L}}^{\dagger} H^{u}_{L} V^{u}_{L} )_{12}= -A^2 \lambda ^5 (i \eta -\rho +1)~.
\eeq

Under the assumption of above rotations, the FCNC operators only arise in left-handed and mixing between first and second generation up-type quarks as
\beq
-{1\over 6}(\bar{u}^{\alpha}_{L}\gamma^{\mu} c^{\alpha}_{L})(\bar{u}^{\beta}_{L}\gamma_{\mu} c^{\beta}_{L})+{1\over 2}(\bar{u}^{\alpha}_{L}\gamma^{\mu} c^{\beta}_{L})(\bar{u}^{\beta}_{L}\gamma_{\mu} c^{\alpha}_{L})~,
\eeq
where the following decomposition satisfied by the color SU(3) fundamental representation has been implemented
\beq
T^{a}_{\alpha\beta} T^{a}_{\gamma\epsilon} = {1\over 2} \delta_{\alpha\epsilon}\delta_{\beta\gamma}-{1\over 6} \delta_{\alpha\beta}\delta_{\gamma\epsilon}~.
\eeq
Under Fierz transformation
\beq
(\bar{u}^{\alpha}_{L}\gamma^{\mu} c^{\beta}_{L})(\bar{u}^{\beta}_{L}\gamma_{\mu} c^{\alpha}_{L})=(\bar{u}^{\alpha}_{L}\gamma^{\mu} c^{\alpha}_{L})(\bar{u}^{\beta}_{L}\gamma_{\mu} c^{\beta}_{L})~,
\eeq
the effective $\Delta C=2$ Hamiltonian can be expressed as
\beq
H^{\Delta C=2}_{AG}=C(\mu)(\bar{u}^{\alpha}_{L}\gamma^{\mu} c^{\alpha}_{L})(\bar{u}^{\beta}_{L}\gamma_{\mu} c^{\beta}_{L})~
\eeq
and the leading order Wilson coefficient at the scale $m_G$ is
\beq
C(m_{G}) = {g^{2} A^4 \lambda ^{10} (1-\rho+i\eta)^2 \over {3 m^{2}_{G}}}~.
\eeq
It means that the $D^{0}-\bar{D}^{0}$ mixing in this axigluon model has $\lambda^{10}$ suppression due to CKM rotation.
The RGE running of the above Wilson coefficient is well known\footnote{The RGE running is actually dependent on $N_f$, the number of active
flavor via $\beta_0=11-2N_f/3$. From the scale $m_G$ down to $\mu_c$, $N_f$ changes correspondingly from $6$ to $4$. But numerically this effect
is small and we will simply take $N_f=5$ in the RGE running.},
\beq
C(\mu_c)=\left ( \frac{\alpha_s(m_G)}{\alpha_s(\mu_c)}\right )^{6/23} C(m_G)~.
\eeq
Notice that the hadronic matrix element of $\Delta C=2$ operator is 
\beq
\langle \bar{D}^0 \vert \bar{u}^{\alpha}_{L}\gamma^{\mu} c^{\alpha}_{L}\bar{u}^{\beta}_{L}\gamma_{\mu} c^{\beta}_{L}  \vert D^0\rangle \equiv
\frac{2}{3}f_D^2 m_D^2 B_D (\mu_c)~.
\eeq
Just like the $B^0-\bar{B}^0$ mixing case, one may define the renormalization group invariant parameter $\hat{B}_D$ by
\beq
\hat{B}_D \equiv (\alpha_s(\mu_c))^{-6/23} B_D(\mu_c)~,
\eeq
which should be $O(1)$. Then the axigluon induced $\Delta C=2$ effective operator contributes to mass difference of neutral D system as
\beq
   \Delta m_D =  \alpha_s(m_G)^{6/23} \frac{8\pi f_D^2 m_D \hat{B}_D}{9}\frac{\alpha_s(m_G)A^4 \lambda ^{10}((1-\rho)^2+\eta^2)}{m^2_G}~.
\eeq

The axigluon model can also induce $\Delta C=1$ effective operator which would in principle affect $D^0-\bar{D}^0$ mixing by  
\begin{align}
\left (M-\frac{i}{2}\Gamma \right )_{12} &= \frac{1}{2m_D} \langle \bar{D}^0 \vert H^{\Delta C=2}_{eff} \vert D^0 \rangle +
  \frac{1}{2m_D} \sum_{n} \frac{\langle \bar{D}^0\vert H^{\Delta C=1}_{eff}\vert n \rangle \langle n \vert H^{\Delta C=1}_{eff} \vert D^0 \rangle}{m_D-E_n+i\epsilon}~.
\end{align}
Actually the experimental observation of comparably large mass and width differences \cite{Nakamura:2010zzi} 
\beq
x\equiv \frac{\Delta m_{D}}{\Gamma_{D}} = 0.98^{+0.24}_{-0.26}\%~, \hspace*{1cm} 
y\equiv \frac{\Delta \Gamma_D}{2\Gamma_D} = (0.83 \pm 0.16)\% 
\eeq
strongly implies that they are dominated by the long distance effects of the SM $\Delta C=1$ operators. Therefore the axigluon induced $\Delta C=1$ terms could be safely neglected as they should be much smaller than the tree-level SM $\Delta C=1$ terms.

Taking the Wolfenstein parameters as \cite{Charles:2004jd}
\beq
   A=0.812~, \hspace*{0.5cm} \lambda=0.2254~,\hspace*{0.5cm} \rho=0.148~,\hspace*{0.5cm} \eta=0.351~
\eeq
and $f_D=207$ MeV \cite{Nakamura:2010zzi}, we obtain
\beq
     \left ( \frac{\Delta m_D}{\Gamma_D}\right )_{axigluon} = 0.082 \% \left ( \frac{1~\mbox{TeV}}{m_G}\right )^2 \left ( \frac{\hat{B}_D}{1}\right )
\eeq
which is roughly one order of magnitude smaller than the experimental result. 

\subsection{Electroweak constrains on the $W'$ model}
\label{sec:ewpt}

In general, the $W'$ must generate its mass through gauge symmetry breaking, then some other neutral component in the $W'$ symmetry breaking sector (for instance $W^3_R$ in the $SU(2)_R$ symmetry breaking) would inevitably mix with $W_L^3$ and some extra $U(1)$ so that the ${W'}^\pm$ would be charged under $U(1)_{em}$. As a consequence, there is a large $Z$-$Z'$ mixing which is constrained by the electroweak precision test. In general, the bound from EWPT is subtle since different fermions $W/Z$ boson couplings are modified in different ways which may even depends on models and a careful global fit is needed. The full results for the $W'$ model to explan the top forward backward asymmetry will be presented elsewhere. Since the overall modification for fermion $Z$ boson coupling is small (except for some right-handed quarks charged under $SU(2)_R$), we only consider the tree-level $Z$-$Z'$ mixing as a rough estimation. For observables that strongly depends on $u/d/b_R$-$Z$ coupling, such as $g_R^2$, $Q_W(Cs)$, etc., their deviations from the SM results are still at the same level as $Z$ mass which is transmitted into $W$ boson mass.

We can start to consider a simple $SU(2)_R \times U(1)_X \times SU(2)_L$ model to estimate how large is the electroweak constraint for the $Z'-Z$ mixing. The $SU(2)_R$ is separated from $SU(2)_L$ to avoid the troublesome $W' - W$ mixing. The two double Higgs field $h_L$ and $h_R$ are charged under $SU(2)_L \times U(1)_X$ and $U(1)_X \times SU(2)_R$ respectively. The higgs fields get their vacuum expectation values $ \langle h_L \rangle = u_L$ and $\langle h_R \rangle = u_R$ which spontaneously break $SU(2)_R \times U(1)_X \times SU(2)_L$ into the diagonal group $U(1)_{em}$. In order to raise the $Z^\prime$ mass so we have less constrain from the $Z$-$Z^\prime$ mixing, we choose $h_R$ transform as a triplet under $SU(2)_R$ so $m_{Z^\prime} = \sqrt{2} m_{W^\prime}$. The gauge quantum number for $h_L$ and $h_R$ are $(0,1, 1/2)$ and $(1,2,0)$ under $SU(2)_R \times U(1)_X \times SU(2)_L$ respectively. For SM fermions, at least the quark doublet $(t,d)_R$ is charged under $SU(2)_R$ (It is possible to have some extra hidden fermions charged under $SU(2)_R$ and $U(1)_X$ to cancel the gauge anomaly). For the rest SM fermions, their quantum number is the same as the SM one if one replace their hyper charge as the $U(1)_X$ charge. The quantum number for $(t,d)_R$ and $(u,b)_R$ under $SU(2)_R \times U(1)_X \times SU(2)_L$ are (1/2, 1/3, 0).

The kinetic term for the link fields $\rm{Tr}[(D_{\mu} h_L)^{\dag} (D_{\mu} h_L)]$ + $\rm{Tr}[(D_{\mu} h_R)^{\dag} (D_{\mu} h_R)]$ becomes the mass terms for the massive gauge bosons. 
The mass matrix of the gauge bosons is  
\begin{eqnarray}
\frac{1}{4}
\begin{pmatrix}
 A^R_{\mu} & A^X_{\mu} & A^L_{\mu}
\end{pmatrix}
\begin{pmatrix}
2 g_R^2 u_R^2 & - 2 g_R g_X u_R^2 & 0 \\
- 2 g_R g_X u_R^2 &  g_X^2 (2 u_R^2 + u_L^2) & -g_X g_L u_L^2 \\
0 & -g_X g_L u_L^2 & g_L^2 u_L^2
\end{pmatrix} 
\begin{pmatrix}
A^R_{\mu} \\ A^X_{\mu} \\ A^L_{\mu}
\end{pmatrix}
 \label{G_MG} \ .
\end{eqnarray}
We introduce the parameter $\epsilon \equiv u_L^2 / 2 u^2_R \ll 1$ which shows that the right-handed symmetry breaking is only a perturbation.  


This matrix can be
diagonalized by means of an orthogonal matrix which we shall call {\bf R}:
\bea
\begin{pmatrix} 
A^R_{\mu} \\ A^X_{\mu} \\ A^L_{\mu} \cr 
\end{pmatrix} = {\rm \bf R^\dagger }
\begin{pmatrix}  A_\mu \cr  Z_\mu \cr  Z^\prime_\mu \cr 
\end{pmatrix} \ ,
\eea
where the mass eigenstates are denoted by $A$, $Z$, and $Z^\prime$. The
eigenstate $A$ is massless and identified as the photon. The couplings
of our theory are related to the electric charge
by
\bea
g_R = \frac{e}{\sin \phi \cos \tw} \, , \quad
g_X = \frac{e}{\cos \phi \cos \tw} \, , \quad
g_L = \frac{e}{\sin \tw}
\eea
where $\tw$ is the weak mixing angle (in the limit $\epsilon \to 0$) and $\phi$ is an additional mixing
angle.  The other two eigenmasses are
\bea
\label{eq:zmass}
m_Z^2 &=& \frac12 u_L^2 (g_Y^2+g_L^2 ) \left[1 - \epsilon  \frac{g_X^4}{(g_R^2+g_X^2)^2} \right]= \frac12 u_L^2 (g_Y^2+g_L^2 ) \left[1 - \epsilon \sin^4 \phi \right] \ ,\\
m_{Z^\prime}^2 &=& \frac12 u_R^2 (g_R^2+g_X^{2}) \left[1 + \epsilon  \frac{g_X^4}{(g_R^2+g_X^2)^2} \right] = \frac12 u_R^2 (g_R^2+g_X^2 ) \left[1 + \epsilon \sin^4 \phi \right] \ ,
\eea
where we have dropped ${\cal O}(\epsilon^2)$ terms and  
\bea
\frac{1}{g_Y^2} \equiv \frac1{g_R^2} + \frac1{g_X^2}\ .
\eea
Clearly, $Z$ is identified with the SM $Z$ boson while $Z^\prime$ is referred to as the heavy $Z$ boson.

\begin{figure}[htbp]
\begin{center}
\includegraphics[angle=0,width=8cm]{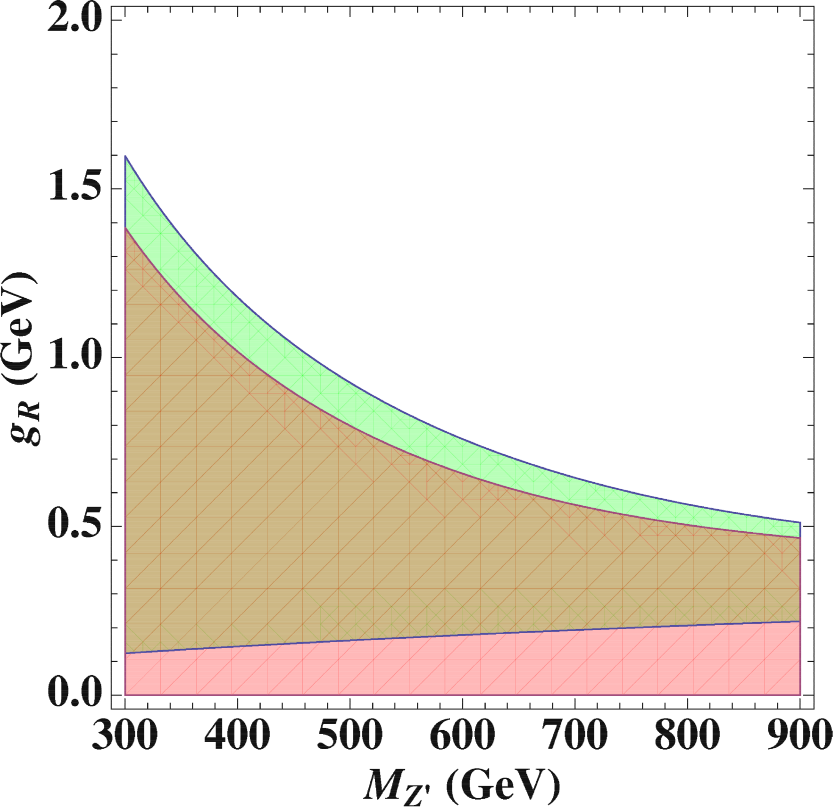}   
\caption{The excluded region on the parameter spaces of $W_R$ model based on tree level contribution to T parameter alone from the new physics at 90$\%$ C. L. The green plus yellow region are is the excluded region for SM Higgs mass $m_h = 114$GeV while the red plus yellow region is the excluded region for SM Higgs mass $m_h = 150$GeV. }
\label{fig:ewpt}
\end{center}
\end{figure}

For small $\epsilon$, the mixing matrix {\bf R} has the following approximate form:
\bea
\footnotesize {\rm \bf R} = \left (
\begin{matrix}
\sin \phi \cos \tw &
\cos \phi \cos \tw & \sin \tw \cr \sin \phi \sin \tw
+ \epsilon \frac{\sin^3 \phi \cos^2 \phi}{\sin \tw} & \cos \phi \sin \tw
- \epsilon \frac{\cos \phi \sin^4 \phi}{\sin \tw} & - \cos \tw \cr
- \cos \phi + \epsilon \cos \phi \sin^4 \phi & \sin \phi
+ \epsilon \cos^2 \phi \sin^3 \phi  & - \epsilon \cot \tw \cos \phi \sin^3
\phi \cr 
\end{matrix}
\right )\ ,
\eea
from which it is simple to derive the SM fermion couplings. The SM fermion and Higgs couplings to $Z$ and $Z^\prime$ can be written as 
\bea
& & g_R \sin \phi \sin \theta_W T_R^3 + g_X \cos \phi \sin \theta_W Q_X - g_L \cos \theta_W T_L^3 \nonumber \\
&=& \frac{e}{\sin \tw \cos \tw} \left( \sin^2 \tw Q - T_L^3 + \epsilon \sin^2 \phi \cos^2 \phi T_R^3 -\epsilon \sin^4 \phi Q_X \right) ,
\eea
\bea
& & g_R (- \cos \phi + \epsilon \cos \phi \sin^4 \phi)T_R^3 + g_X (\sin \phi + \epsilon \cos^2 \phi \sin^3 \phi)Q_X - g_L (- \epsilon \cot \tw \cos \phi \sin^3 \phi)T_L^3 \nonumber \\
&=& \frac{e}{\cos \tw} \left( - \frac{\cos \phi}{\sin \phi} T_R^3 + \frac{\sin \phi}{\cos \phi} Q_X + \frac{ \epsilon \cos \phi \sin^3 \phi}{\sin^2 \theta_W} ( -T_L^3 +Q \sin^2 \theta_W) \right) .
\eea
In the limit of large $SU(2)_R$ breaking vev ($\epsilon \ll 1$) and small mixings ($\phi \rightarrow 0$), the Higgs current can be approximated as (we drop the $\epsilon \cos^2 \phi \sin^3 \phi$ term) 
\bea
J_{Z^\prime}^\mu (h) = -\frac{1}{2} g_X \sin \phi (h^\dag D^\mu h) + h . c. \ ,
\eea
which induce a dimension six operator
\bea
a_h \mathcal{O}_h = - \frac{g_R^2+g_X^2} {2 m_{Z^\prime}^2} \sin^4 \phi (h^\dag D^\mu h)^2 \ .
\eea 
which coincident with Eq. (\ref{eq:zmass}) that $\Delta m_Z^2 = - \epsilon \sin \phi^4 m_Z^2$.  
We can calculate the corresponding T parameter from the tree level gauge boson mixing, 
\bea
T = - \frac{a_h v^2}{\alpha_f} = \frac{ \epsilon \sin^4 \phi}{\alpha_f}  
\eea 

Using the SM model $m_Z$, $G_F$ (the life time of $\tau$) and $\alpha \equiv e^2 / 4 \pi$ as the basic input parameter, we can calculate the allowed parameter space including the Higgs radiative corrections according to the most recent results: $S = 0.03 \pm 0.09$ and $T = 0.07 \pm 0.08$ (with 87$\%$ strong correlation) \cite{Nakamura:2010zzi}. The results are presented in Fig \ref{fig:ewpt}. We can see that for sufficient heavy $Z'$ and strong coupling $g_R$ (for instance, $g_R =2$, $m_{Z'} = 900$ GeV which is used in Ref. \cite{kathryn}), it is well above the excluded region.   

\section{Conclusions}
\label{sec:conclusion}
We discuss the feature of various models 
for the top quark forward-backward asymmetry anomaly at
Tevatron, using the latest CDF measurements on total asymmetry in
lab frame $A^{/rm lab}_{FB}$, the rapidity dependent asymmetry
$A_{FB}(\mid \Delta \eta\mid>1)$, the mass depedent asymmetry 
$A_{FB}(M_{t\bar{t}}>450~\text{GeV})$ and the total $t\bar{t}$
production cross section $\sigma_{t\bar{t}}$. 

The axigluon model in \cite{paul} has difficulty in explain the large rapidity 
dependent asymmetry and the mass dependent asymmetry
simultaneously. In addition, the latest dijet search \cite{dijet} at ATLAS 
has ruled out the parameter region that is relevant to top $A_{FB}$. On the other hand, in contrast to the conclusion in Ref. \cite{cp}, the model itself does not suffer from the flavor $B_d$ mixing under flavor protection $U(1)_d$ and a careful calculation shows that their to $D^0-\bar{D}^0$ is still one order lower than the current experimental bound.

The $t$-channel $Z^{\prime}$ \cite{hitoshi}, $W^{\prime}$\cite{waiyee} 
and anti-triplet diquark \cite{tim} models all have parameter 
regions that satisfy all three asymmetry
measurements within 1~$\sigma$. However, the corresponding production cross section predicted by
the 1~$\sigma$ asymmetry requirement in the $Z^{\prime}$ 
model are always significantly below the 1~$\sigma$ of cross section measurement. 
The best fit point of $Z^{\prime}$ is about 700 GeV with purely
righthanded coupling $g^R_{ut}\simeq 1.8$ which
corresponds to 6.9 pb.
However, this best fit point will generate a large number of 
same-sign top quark events at Tevatron which is at least five times larger than the SM prediction. 
We conclude that the $Z^{\prime}$ model is very difficult to
be consistent with all the measurements.

Both $W^{\prime}$ and anti-triplet diquark models 
predict the cross sections are larger than the measurement
but various factors can lower the survival efficiency after cuts in these models
to ease the tension between asymmetry and cross section.
The best fit point for anti-triplet diquark lies in very high mass region
and with better survival efficiency \cite{kathryn}, it is difficult to
fit the differential cross section $d\sigma/d M_{t\bar{t}}$. A rough estimation for $W'$ model shows that the bounds from electroweak precision tests are weak due the heavy $Z'$ and strongly coupling $g_R$. Therefore, we conclude that the best model is the $t$-channel $W^{\prime}$ model at the current stage. To test such model directly, we also use the 1~$\sigma$ asymmetry parameters to compute the production rate of $t\bar{t}+j$ from $tW^{\prime}$ at 7 TeV LHC and 
the production rate is 10\%-50\% of SM $t\bar{t}$.

Last, we want to mention that the latest NNLL calculation $\sigma_{t\bar{t}}(m_{t}=173.1~\text{GeV})= 6.30\pm 0.19^{+0.31}_{-0.23}~\text{pb}$ \cite{qcd} is significantly lower than the experimental results. If the result does not significantly change for $m_{t}=172.5~\text{GeV}$ which is used for Tevatron experiments, then the fits for t-channel $W'$ and anti-triplet would be better while the t-channel $Z'$ would be worse. 

\section*{Acknowledgement}
J.S. and K.W would like to thank Zhejiang Institute for Modern Physics at Zhejiang University and Prof. Mingxing Luo for hospitality after the Tohuku earthquake. We would like to thank Qinghong Cao, Mingxing Luo, Hitoshi Murayama, David Shih, Matt Strassler, Scott Thomas and Carlos Wagner for useful discussion. We also thank Tim Tait who initiate the electroweak bounds for $W'$ model from our discussion. The work is partially supported by the World Premier International Research Center Initiative (WPI initiative) MEXT, Japan.  J.S. and K.W. are also supported by the Grant-in-Aid for scientific research (Young Scientists (B) 21740169) and (Young Scientists (B) 22740143) from Japan Society for the Promotion of Science (JSPS), respectively. G.Z is supported in part by the National Science Foundation of China (No. 11075139 and No.10705024) and the Fundamental Research Funds for the Central Universities.

\section*{Notes added}
While this work was being delayed by the huge earthquake in Japan, Ref. \cite{kathryn} appeared, which overlap with ours in the study of fitting different models on different top forward backward asymmetries. Our results agree with quantitatively with theirs for fitting different top forward backward asymmetries and the total $t \bar{t}$ cross section. However, we notice that the axigluon model and the heavy $Z'$ model are severely constrained by the dijet search at the LHC (ATLAS) and the same sign dilepton search at the Tevatron (CDF). Therefore, we conclude that the heavy $W'$ model is the most promising one at present. We also consider the indirect bounds for different models.

\end{document}